\documentclass[10pt]{article}
\pdfoutput=1
\usepackage{chicago,epic,graphpap,graphics,float,tabularx,paralist,longtable,fancyhdr,fancyvrb,bbm}
\usepackage{amsmath}
\usepackage{amstext}
\usepackage{amsbsy}
\usepackage{amsthm}
\usepackage{amsfonts}
\usepackage{amssymb}
\usepackage{amscd}
\usepackage{eucal}
\def\S{{\bf S}}
\def\m{\mu}
\def\d{\delta}
\def\D{\Delta}
\def\t{\tau}
\def\p{\pi}
\def\th{\theta}
\def\g{\gamma}

\def\as{\quad\text{{\rm a.s.}}}

\def\1{{\mathbbm 1}}

\def\P{{\mathbb P}}
\def\R{{\mathbb R}}
\def\N{{\mathbb N}}
\def\eqdef{\triangleq}
\def\intt{\int_0^t}

\def\F{{\cal F}}
\def\ph{\varphi}
\def\half{\frac{1}{2}}
\def\brac#1{\langle #1 \rangle}

\begin{document}

\centerline{\Large\bf Variations on an example of Karatzas and Ruf}
\vspace{15pt} \centerline{\large Robert Fernholz\footnote{INTECH, One Palmer Square, Princeton, NJ 08542.  bob@bobfernholz.com. The author thanks Ren\'e Carmona, Ioannis Karatzas, and Johannes Ruf for their invaluable comments and suggestions regarding this research.}} 
\centerline{\today}
\vspace{15pt}
\begin{abstract}
Markets composed of stocks with capitalization processes represented by positive continuous semimartingales are studied under the condition that the market excess growth rate is bounded away from zero. The following examples of these markets are given: {\em i})~a market with a singular covariance matrix and instantaneous relative arbitrage; {\em ii})~a market with a singular covariance matrix and no arbitrage; {\em iii})~a market with a nonsingular covariance matrix and no arbitrage; {\em iv})~a market with a nonsingular covariance matrix and relative arbitrage over an arbitrary time horizon.
\end{abstract}
\vspace{15pt}

For $n\in\N$ and $T\in(0,\infty)$, consider a market composed of stocks with capitalization processes $X_1,\ldots,X_n$ represented by positive continuous semimartingales defined on $[0,T]$.  \citeN{KR:2015} give an example of such a market in which  the capitalization processes are martingales, the market covariance matrix is singular,  and the excess growth rate $\g^*_\m$ of the market portfolio is bounded away from zero.   The condition that the capitalization processes are martingales is of interest because arbitrage is not possible in a market with martingale capitalization processes (see, e.g., \citeN{Karatzas/Shreve:1998}). The condition that $\g^*_\m$ is bounded away from zero is of interest because this implies the existence of relative arbitrage over long enough time horizons (see \citeN{FK:2005}). 

Here we present five variations on the example of  \citeN{KR:2015}. The first variation is an example of a market with a singular covariance matrix and instantaneous relative arbitrage; the second is a market with a singular covariance matrix and martingale capitalization processes; the third is a market with a nonsingular covariance matrix and martingale capitalization processes; and the fourth is a market with a nonsingular covariance matrix and relative arbitrage over an arbitrary time horizon. In these four examples, the market capitalization process is a martingale; in the fifth example this condition is relaxed. 

A market is {\em strongly nondegenerate} if the eigenvalues of its covariance matrix are  bounded away from zero. 
In all the examples we consider here the markets are {\em diverse}, i.e., the market weights are all bounded away from one. In a diverse, strongly nondegenerate market, relative arbitrage exists over an arbitrary time horizon (see Section~8 of \citeN{FK:2009}), so the capitalization processes cannot be martingales.  Hence, in our third example, nonsingularity of the market covariance matrix cannot be strengthened to strong nondegeneracy, since arbitrage is not possible in a market with martingale capitalization processes. 

\vspace{10pt}
\noindent{\bf Definition 1.} For  $T>0$, a market defined on $[0,T]$ has {\em relative arbitrage} if there exist portfolios $\nu$ and $\eta$ with value processes $Z_\nu$ and $Z_\eta$ such that 
\begin{align*}
\P\big[Z_\nu(T)/Z_\eta(T)&\ge Z_\nu(0)/Z_\eta(0)\big]=1,\\
\P\big[Z_\nu(T)/Z_\eta(T)&> Z_\nu(0)/Z_\eta(0)\big]>0.
\end{align*}

\vspace{5pt}
\noindent{\bf Definition 2.} For  $T>0$, a market defined on $[0,T]$ has  {\em instantaneous relative arbitrage} if there exist portfolios $\nu$ and $\eta$ with value processes $Z_\nu$ and $Z_\eta$ such that for any $t_1,t_2\in[0,T]$ with $t_1<t_2$,
\[
\P\big[Z_\nu(t_2)/Z_\eta(t_2)> Z_\nu(t_1)/Z_\eta(t_1)\big]=1.
\]

\pagebreak
\noindent{\bf Example 1.}
\vspace{5pt}

Here we give an example of a market with $\g^*_\m$ bounded away from zero and a singular covariance matrix. In this example the market weight processes are confined to a circle, and this generates instantaneous relative arbitrage.

Let $a\in(0,1/3)$ be a real constant, and consider the time horizon $[0,T]$. Let $(W,\th)$ be a 2-dimensional Brownian motion with the usual filtration $\F$, and let $(X_1,X_2,X_3)$ be the market defined  by
\[
X_i(t)=e^{W(t)-t/2}\Big(\frac{1}{3}+a\cos\big(\th(t)+(i-1)2\p/3\big)\Big),\qquad 0\le t\le T,
\]
for $i=1,2,3$.  The market capitalization process $X$ is given by
\begin{align*}
X(t)&=X_1(t)+X_2(t)+X_3(t)\\
&= e^{W(t)-t/2},
\end{align*}
so the market weights are
\begin{equation}\label{0}
\m_i(t)=\frac{X_i(t)}{X(t)}=\frac{1}{3}+a\cos\big(\th(t)+(i-1)2\p/3\big),
\end{equation}
for $i=1,2,3$. Hence, $0<\m_i(t)<2/3$, for $i=1,2,3$,  and the market is diverse. Since for all $x\in\R$,
\begin{equation}\label{00}
\sum_{i=1}^3\sin^2\big(x+(i-1)2\p/3\big) = \sum_{i=1}^3\cos^2\big(x+(i-1)2\p/3\big) = \frac{3}{2}.
\end{equation}
we  have
\begin{equation}\label{1}
\m_1^2(t)+\m_2^2(t)+\m_3^2(t)=\frac{1}{3}+\frac{3a^2}{2}<\half,
\end{equation}
so the points $(\m_1(t),\m_2(t),\m_3(t))\in\R^3$ lie on the intersection of the plane
\[
x_1+x_2+x_3=1
\]
with the sphere of radius $\sqrt{1/3+3a^2/2}$ centered at the origin. This intersection is a circle of radius $\sqrt{3a^2/2}<\sqrt{1/6}$
centered at $(1/3,1/3,1/3)$, and this circle lies in the  (open) simplex 
\[ 
\D^3 \eqdef \big\{x\in\R^3: x_1+x_2+x_3 = 1, x_i>0\big\}.
\]

From \eqref{0} we see that for $i=1,2,3$,
\[
d\m_i(t)= -a\sin\big(\th(t)+(i-1)2\p/3\big)d\th(t) - \frac{a}{2} \cos\big(\th(t)+(i-1)2\p/3\big)dt,\as,
\]
so
\[
d\brac{\m_i}_t = a^2\sin^2\big(\th(t)+(i-1)2\p/3\big)dt,\as,
\]
and
\begin{align}
\t_{ii}(t)&=\frac{1}{\m^2_i(t)}\frac{d\brac{\m_i}_t}{dt}\notag\\
&= \frac{a^2\sin^2\big(\th(t)+(i-1)2\p/3\big)}{\m^2_i(t)},\as\label{2}
\end{align}

Consider the portfolio generating function 
\[
\S(x)=\big(x_1^2+x_2^2+x_3^2\big)^{1/2}.
\]
This function generates the portfolio $\p$ with weights
\begin{equation}\label{2.1}
\p_i(t)=\frac{\m_i^2(t)}{\m_1^2(t)+\m_2^2(t)+\m_3^2(t)}=\frac{\m_i^2(t)}{1/3+3a^2/2}
\end{equation}
and with value function $Z_\p$ that satisfies
\[
d\log\big(Z_\p(t)/Z_\m(t)\big)=d\log \S(\m(t))-\g^*_\p(t)\,dt,\as,
\]
where $Z_\m$ is the market value process (see \citeN{F:2002}, Example~3.1.9). Since \eqref{1} implies that $\S(\m(t))$ is constant, this reduces to
\begin{equation}\label{3}
d\log\big(Z_\p(t)/Z_\m(t)\big)=-\g^*_\p(t)\,dt,\as
\end{equation}
Since this has no stochastic component, the relative variance $\t_{\p\p}(t)$ vanishes for all $t\in[0,T]$, so 
\begin{align}
\g^*_\p(t) &= \half\Big(\sum_{i=1}^3 \p_i(t)\t_{ii}(t)-\t_{\p\p}(t)\Big)\notag\\
&= \half\sum_{i=1}^3 \p_i(t)\t_{ii}(t)\notag\\
&= \frac{a^2}{2/3+3a^2}\sum_{i=1}^3\sin^2\big(\th(t)+(i-1)2\p/3\big)\notag\\
&= \frac{3a^2}{4/3+6a^2}>0,\label{33}
\end{align}
by \eqref{00}, \eqref{2}, and \eqref{2.1}. From \eqref{3}  and \eqref{33} it follows that in this market there is instantaneous relative arbitrage.

On the other hand, from \eqref{2} we have
\begin{align*}
\m_i(t)\t_{ii}(t)&=\frac{a^2\sin^2\big(\th(t)+(i-1)2\p/3\big)}{\m_i(t)}\\
&> \frac{3a^2\sin^2\big(\th(t)+(i-1)2\p/3\big)}{2},\as,
\end{align*}
for $i=1,2,3$, since $0<\m_i(t)<2/3$, so \eqref{00} implies that
\[
\g^*_\m(t)=\half\sum_{i=1}^3\m_i(t)\t_{ii}(t)> \frac{9a^2}{8},\as,
\]
for  $t\in[0,T]$. 

Of the two processes that drive this market, $\th$ generates circular motion in the plane of $\D^3$ about the point $(1/3,1/3,1/3)\in\R^3$,  and $W$  generates radial motion from the origin in the positive orthant. The motion due to $W$ does not lie in the plane of $\D^3$, so the rank of the market covariance matrix will be two. 

In the next example, we shall show that the same conditions as in this example, i.e., $\g^*_\m$ bounded away from zero and market covariance matrix of rank two, can result in a market with no arbitrage.

\vspace{10pt}
\noindent{\bf Example 2.}
\vspace{5pt}

Here we give an example of a  market with $\g^*_\m$ bounded away from zero, a singular covariance matrix, and martingale capitalization processes. In this example the market weight processes are confined to an expanding circle, similarly to \citeN{KR:2015}, in which the market weight processes are confined to an expanding annulus. In this example there is no arbitrage, since arbitrage is not possible in a market with martingale capitalization processes.

Let $a\in(0,1/3)$ be a real constant and consider the time horizon $[0,T]$ with $T<-2\log (3a)$.  Let  $(W,\th)$ be a 2-dimensional Brownian motion with the usual filtration $\F$, and let $(X_1,X_2,X_3)$ be the market defined by
\[
X_i(t)=e^{W(t)-t/2}\Big(\frac{1}{3}+ae^{t/2}\cos\big(\th(t)+(i-1)2\p/3\big)\Big),\qquad 0\le t\le T,
\]
for $i=1,2,3$. In this case,
$e^{W(t)-t/2}$ and $e^{t/2}\cos\big(\th(t)+(i-1)2\p/3\big)$ are independent $\F$-martingales, so the $X_i$ are also $\F$-martingales. As a result, arbitrage cannot exist in to this market.

The market capitalization process $X$ is given by
\begin{align*}
X(t)&=X_1(t)+X_2(t)+X_3(t)\\
&= e^{W(t)-t/2},
\end{align*}
so the market weights are
\[
\m_i(t)=\frac{X_i(t)}{X(t)}=\frac{1}{3}+ae^{t/2}\cos\big(\th(t)+(i-1)2\p/3\big),
\]
for $i=1,2,3$. In this case,
\[
d\m_i(t)=- ae^{t/2}\sin\big(\th(t)+(i-1)2\p/3\big)d\th(t),\as,
\]
so the $\m_i$ are $\F$-martingales. 

As in Example~1, we have $0<\m_i(t)\le 1/3+ae^{t/2}<2/3$, for $i=1,2,3$, so the market is diverse. It follows from \eqref{00} that
\[
\m_1^2(t)+\m_2^2(t)+\m_3^2(t)=\frac{1}{3}+\frac{3a^2 e^t}{2}<\half,\qquad 0\le t\le T,
\]
so the points $(\m_1(t),\m_2(t),\m_3(t))\in\R^3$ lie on the intersection of the plane
\[
x_1+x_2+x_3=1
\]
with the sphere of radius $\sqrt{1/3+3a^2e^t/2}$ centered at the origin. This intersection is a circle of radius  $\sqrt{3a^2e^t/2}<\sqrt{1/6}$ centered at $(1/3,1/3,1/3)$, and this circle lies in the  simplex $\D^3$.

Similarly to \eqref{2}, for $i=1,2,3$, we have
\begin{equation}\label{4}
\t_{ii}(t) = \frac{a^2e^t\sin^2\big(\th(t)+(i-1)2\p/3\big)}{\m^2_i(t)},\as,
\end{equation}
so
\begin{align*}
\m_i(t)\t_{ii}(t)&=\frac{a^2e^t\sin^2\big(\th(t)+(i-1)2\p/3\big)}{\m_i(t)}\\
&> \frac{3a^2e^t\sin^2\big(\th(t)+(i-1)2\p/3\big)}{2},\as,
\end{align*}
since $0<\m_i(t)<2/3$. Hence, it follows from \eqref{00} that
\[
\g^*_\m(t)=\half\sum_{i=1}^3\m_i(t)\t_{ii}(t)> \frac{9a^2e^t}{8}\ge \frac{9a^2}{8},\as,
\]
for  $t\in[0,T]$.  

Of the two martingales that drive this market, $\th$ generates circular motion in the plane of $\D^3$ about the point $(1/3,1/3,1/3)\in\R^3$,  and $W$  generates radial motion from the origin in the positive orthant. The motion due to $W$ does not lie in the plane of $\D^3$, so the rank of the market covariance matrix will be two. 

In the next example, we shall perturb the current model radially in the plane of $\D^3$, and this will result in a nonsingular market covariance matrix. This perturbation will not alter the martingale structure of the capitalization processes, so the resulting market will still not permit arbitrage.

\pagebreak
\noindent{\bf Example 3.}
\vspace{5pt}

 Here we give an example of a market with $\g^*_\m$ bounded away from zero, a nonsingular covariance matrix, and martingale capitalization processes. In this example the market weight processes are confined to an expanding annulus, as in \citeN{KR:2015}, where the market covariance matrix was singular. Although the market covariance matrix is nonsingular in this example, the market cannot be strongly nondegenerate, since strong nondegeneracy in a diverse market implies the existence of relative arbitrage over an arbitrary time horizon (see Section~8 of \citeN{FK:2009}), and arbitrage is not possible with martingale capitalization processes.

Let $a\in(0,1/9)$ be a real constant and consider the time horizon $[0,T]$ with $T<-2\log(9a)$. Let $(W,\th,B)$ be a 3-dimensional Brownian motion with the usual filtration $\F$, and let $(X_1,X_2,X_3)$ be the market defined  by
\[
X_i(t)=e^{W(t)-t/2}\Big(\frac{1}{3}+\ph(t)e^{t/2}\cos\big(\th(t)+(i-1)2\p/3\big)\Big),\qquad 0\le t\le T,
\]
for $i=1,2,3$, where $\ph$ is a  continuous $\F$-martingale driven by $B$ such that $a<\ph(t)<3a$. We shall establish the precise structure  of the process $\ph$ below. Since $e^{W(t)-t/2}$, $e^{t/2}\cos\big(\th(t)+(i-1)2\p/3\big)$, and $\ph$ are all independent $\F$-martingales, the $X_i$ will also be $\F$-martingales.  As a result, arbitrage cannot exist in to this market.

The market capitalization process $X$ is given by
\begin{align*}
X(t)&=X_1(t)+X_2(t)+X_3(t)\\
&= e^{W(t)-t/2},
\end{align*}
and the market weights are
\[
\m_i(t)=\frac{X_i(t)}{X(t)}=\frac{1}{3}+\ph(t)e^{t/2}\cos\big(\th(t)+(i-1)2\p/3\big),
\]
for $i=1,2,3$. In this case,
\[
d\m_i(t)=- \ph(t)e^{t/2}\sin\big(\th(t)+(i-1)2\p/3\big)d\th(t)+e^{t/2}\cos\big(\th(t)+(i-1)2\p/3\big)d\ph(t),\as
\]
so the $\m_i$ are all $\F$-martingales. 

As in Examples~1 and~2, we have $0<\m_i(t)< 1/3+3ae^{t/2}<2/3$, a.s., for $i=1,2,3$, so the market is a.s.\ diverse. It follows from \eqref{00} that
\[
\m_1^2(t)+\m_2^2(t)+\m_3^2(t)=\frac{1}{3}+\frac{3\ph^2(t) e^t}{2}<\half,\as,\qquad 0\le t\le T,
\]
so the points $(\m_1(t),\m_2(t),\m_3(t))\in\R^3$ lie on the intersection of the plane
\[
x_1+x_2+x_3=1
\]
with the sphere of radius $\sqrt{1/3+3\ph^2(t)e^t/2}$ centered at the origin. This intersection is a circle of radius $\sqrt{3\ph^2(t)e^t/2}<\sqrt{1/6}$ centered at $(1/3,1/3,1/3)$, and this circle lies in the  simplex $\D^3$.  

Now, define the process $\ph$ by 
\begin{equation}\label{4.1}
\ph(t)=2a+\psi(t),
\end{equation}
where 
\begin{equation}\label{4.2}
\psi(t)=\intt \big(a^2-\psi^2(s)\big)dB(s),
\end{equation}
for $t\in[0,T]$,  and $B$ is the Brownian motion introduced above. In this case,  $\psi$ is a continuous $\F$-martingale and
\begin{equation}\label{5}
\P\big[-a<\psi(t)<a, t\in[0,T]\big]=1.
\end{equation}
The process $\psi$ was suggested by Ioannis Karatzas. The structural details of this process can be verified in \citeN{Karatzas/shreve:1991}, Proposition 5.5.22(d) and Theorem 5.5.29. This process $\psi$ is a linear diffusion with state space $(-a,a)$ and visits, with positive probability, any given neighborhood $U\subset (-a,a)$ during any time interval $(0,\d]$, for $\d>0$ (see Theorem 4.8 of \citeN{KR:2016}, and \citeN{BR:2015}). It follows that  $\ph$ is also a continuous $\F$-martingale with  $a<\ph(t)<3a$, a.s., and 
\begin{equation}\label{6}
d\brac{\ph}_t=d\brac{\psi}_t= \big(a^2-\psi^2(t)\big)^2dt,\as
\end{equation}

Similarly to \eqref{2} and \eqref{4}, we have for $i=1,2,3$,
\begin{equation}\label{5.1}
\t_{ii}(t) = \frac{\ph^2(t)e^t\sin^2\big(\th(t)+(i-1)2\p/3\big)+e^t\cos^2\big(\th(t)+(i-1)2\p/3\big)\big(a^2-\psi^2(t)\big)^2}{\m^2_i(t)},\as,
\end{equation}
so
\begin{align*}
\m_i(t)\t_{ii}(t)&=\frac{\ph^2(t)e^t\sin^2\big(\th(t)+(i-1)2\p/3\big)+e^t\cos^2\big(\th(t)+(i-1)2\p/3\big)\big(a^2-\psi^2(t)\big)^2}{\m_i(t)}\\
&> \frac{3\ph^2(t)e^t\sin^2\big(\th(t)+(i-1)2\p/3\big)+3e^t\cos^2\big(\th(t)+(i-1)2\p/3\big)\big(a^2-\psi^2(t)\big)^2}{2},\as,
\end{align*}
since $0<\m_i(t)<2/3$. Hence, from \eqref{00} we have
\[
\g^*_\m(t)=\half\sum_{i=1}^3\m_i(t)\t_{ii}(t)> \frac{9\ph^2(t)e^t+9e^t\big(a^2-\psi^2(t)\big)^2}{8}> \frac{9a^2}{8},\as,
\]
for  $t\in[0,T]$. 

Of the three martingales that drive this market, $\th$ generates circular motion in the plane of $\D^3$ about the point $(1/3,1/3,1/3)\in\R^3$, $\ph$ generates radial motion from the point $(1/3,1/3,1/3)$ in the plane of $\D^3$, and $W$ generates radial motion from the origin in the positive orthant. It follows from \eqref{5} and \eqref{6} that $d\brac{\ph}_t/dt>0$, a.s., so these three movements span $\R^3$, and the market covariance matrix will be nonsingular. However, since  $\big(a^2-\psi^2(t)\big)$ can be arbitrarily small, the market will not be strongly nondegenerate. 

In the next example, we shall show that the same conditions as in this example, i.e., $\g^*_\m$ bounded away from zero and nonsingular market covariance matrix, can result in a market with relative arbitrage over an arbitrary time horizon.

\vspace{10pt}
\noindent{\bf Example 4.}
\vspace{5pt}

Here we give an example of a market with $\g^*_\m$ bounded away from zero and a nonsingular covariance matrix. In this example the market weight processes range throughout a stationary annulus, and this time homogeneity generates  relative arbitrage over an arbitrary time horizon.

Let $a\in(0,1/9)$ be a real constant and consider the time horizon $[0,T]$. Let $(W,\th,B)$ be a 3-dimensional Brownian motion with the usual filtration $\F$, and let $(X_1,X_2,X_3)$ be the market defined  by
\[
X_i(t)=e^{W(t)-t/2}\Big(\frac{1}{3}+\ph(t)\cos\big(\th(t)+(i-1)2\p/3\big)\Big),\qquad 0\le t\le T,
\]
for $i=1,2,3$, where $\ph$ is the martingale defined in \eqref{4.1} and \eqref{4.2}, with $a<\ph(t)< 3a$, a.s.
 
The market capitalization process $X$ is given by
\begin{align*}
X(t)&=X_1(t)+X_2(t)+X_3(t)\\
&= e^{W(t)-t/2},
\end{align*}
and the market weights are
\begin{equation}\label{12}
\m_i(t)=\frac{X_i(t)}{X(t)}=\frac{1}{3}+\ph(t)\cos\big(\th(t)+(i-1)2\p/3\big),
\end{equation}
for $i=1,2,3$. In this case,
\begin{multline*}
d\m_i(t)=- \ph(t)\sin\big(\th(t)+(i-1)2\p/3\big)d\th(t)\\+\cos\big(\th(t)+(i-1)2\p/3\big)d\ph(t)-\frac{\ph(t)}{2}\sin\big(\th(t)+(i-1)2\p/3\big)dt,\as
\end{multline*}

As in Examples~1, 2, and~3, we have $0<\m_i(t)<1/3+3a<2/3$, a.s., for $i=1,2,3$, so the market is a.s.\ diverse. It follows from \eqref{00} that
\[
\m_1^2(t)+\m_2^2(t)+\m_3^2(t)=\frac{1}{3}+\frac{3\ph^2(t) }{2},\qquad 0\le t\le T,
\]
so
\[
\frac{1}{3}+\frac{3a^2}{2}< \m_1^2(t)+\m_2^2(t)+\m_3^2(t)<\frac{1}{3}+\frac{27a^2}{2}<\half,\as,
\]
so the points $(\m_1(t),\m_2(t),\m_3(t))\in\R^3$ lie on the plane
\[
x_1+x_2+x_3=1
\]
between  the spheres of radius $\sqrt{1/3+3a^2/2}$ and $\sqrt{1/3+27a^2/2}$ centered at the origin. This set is the open annulus 
\begin{equation}\label{13}
A=\big\{(x_1,x_2,x_3)\in\D^3: 3a^2/2<(x_1-1/3)^2+(x_2-1/3)^2+(x_3-1/3)^2<27a^2/2\big\},
\end{equation}
which lies between the circles of radius $\sqrt{3a^2/2}$ and $3\sqrt{3a^2/2}<\sqrt{1/6}$ centered at $(1/3,1/3,1/3)$, both of which lie within the simplex $\D^3$.

Similarly to \eqref{5.1}, we have for $i=1,2,3$,
\[
\t_{ii}(t) = \frac{\ph^2(t)\sin^2\big(\th(t)+(i-1)2\p/3\big)+\cos^2\big(\th(t)+(i-1)2\p/3\big)\big(a^2-\psi^2(t)\big)^2}{\m^2_i(t)},\as,
\]
so
\begin{align*}
\m_i(t)\t_{ii}(t)&=\frac{\ph^2(t)\sin^2\big(\th(t)+(i-1)2\p/3\big)+\cos^2\big(\th(t)+(i-1)2\p/3\big)\big(a^2-\psi^2(t)\big)^2}{\m_i(t)}\\
&> \frac{3\ph^2(t)\sin^2\big(\th(t)+(i-1)2\p/3\big)+3\cos^2\big(\th(t)+(i-1)2\p/3\big)\big(a^2-\psi^2(t)\big)^2}{2},\as,
\end{align*}
since $0<\m_i(t)<2/3$. Hence, from \eqref{00} we have
\[
\g^*_\m(t)=\half\sum_{i=1}^3\m_i(t)\t_{ii}(t)> \frac{9\ph^2(t)+9\big(a^2-\psi^2(t)\big)^2}{8}> \frac{9a^2}{8},\as,
\]
for  $t\in[0,T]$. 

Of the three processes that drive this market, $\th$ generates circular motion in the plane of $\D^3$ about the point $(1/3,1/3,1/3)\in\R^3$, $\ph$ generates radial motion from the point $(1/3,1/3,1/3)$ in the plane of $\D^3$, and $W$ generates radial motion from the origin in the positive orthant. It follows from \eqref{5} and \eqref{6} that $d\brac{\ph}_t/dt>0$, a.s., so these three movements span $\R^3$, and the market covariance matrix will be nonsingular. However, since  $\big(a^2-\psi^2(t)\big)$ can be arbitrarily small, the market will not be strongly nondegenerate.

The structure of the market weights given by \eqref{12} results in a type of time homogeneity for $t>0$, and this produces relative arbitrage over the (arbitrary) time horizon $[0,T]$. If $A\subset\D^3$ is the annulus defined above in \eqref{13}, then for any $t\in(0,T]$,
\[
\P\big[\m(t)\in A\big]=1,
\]
and for any open subset $U\subset A$, 
\[
\P\big[\m(t)\in U\big]>0
\]
(see Theorem 4.8 of \citeN{KR:2016}, and \citeN{BR:2015}). Hence, for any portfolio generating function $\S$ and for any $t\in(0,T]$,
\[
\text{ess inf}\big\{\S(\m(t))\big\}=\inf\big\{\S(x):x\in A\big\}.
\]
This implies that the essential infimum of $\S(\m(t))$ is invariant over $(0,T]$, so it follows from Proposition~1 of \citeN{F:2015b} that relative arbitrage exists in this market.

\vspace{10pt}
\noindent{\bf Example 5.}
\vspace{5pt}

The behavior of the market in Examples~1 through~4 does not depend of the fact that the process $e^{W(t)-t/2}$ is a martingale, since the relevant structure is determined by the market weights alone. Indeed, the market capitalization process $e^{W(t)-t/2}$ could be replaced by any positive continuous semimartingale $\kappa$ that is independent of $\ph$ and $\th$, and the market weights $\m_i$ will remain unchanged. Consider, for instance,  Example~3, where
\[
X_i(t)=e^{W(t)-t/2}\Big(\frac{1}{3}+\ph(t)e^{t/2}\cos\big(\th(t)+(i-1)2\p/3\big)\Big),\qquad 0\le t\le T.
\]
In this case, the market model would become
\[
X_i(t)=\kappa(t)\Big(\frac{1}{3}+\ph(t)e^{t/2}\cos\big(\th(t)+(i-1)2\p/3\big)\Big),\qquad 0\le t\le T,
\]
for $i=1,2,3$. Of course, the filtration $\F$ would have to be adjusted accordingly, but after that, all the analysis would remain the same. In the simplest case, the process $\kappa$, which represents the total capitalization of the market, could be set identically equal to one, and the capitalization processes $X_i$ would be the same as the market weight processes $\m_i$. This can be done in all four of the previous examples.

\bibliographystyle{chicago}
\bibliography{math}

\begin{thebibliography}{}

\bibitem[\protect\citeauthoryear{Bruggeman and Ruf}{Bruggeman and
  Ruf}{2015}]{BR:2015}
Bruggeman, C. and J.~Ruf (2015).
\newblock A one-dimensional diffusion hits points fast.
\newblock {\em ArXiv e-prints\/}.

\bibitem[\protect\citeauthoryear{Fernholz}{Fernholz}{2002}]{F:2002}
Fernholz, R. (2002).
\newblock {\em Stochastic Portfolio Theory}.
\newblock New York: Springer-Verlag.

\bibitem[\protect\citeauthoryear{{Fernholz}}{{Fernholz}}{2015}]{F:2015b}
{Fernholz}, R. (2015, October).
\newblock {An example of short-term relative arbitrage}.
\newblock {\em ArXiv e-prints\/}.

\bibitem[\protect\citeauthoryear{Fernholz and Karatzas}{Fernholz and
  Karatzas}{2005}]{FK:2005}
Fernholz, R. and I.~Karatzas (2005).
\newblock Relative arbitrage in volatility-stabilized markets.
\newblock {\em Annals of Finance\/}~{\em 1}, 149--177.

\bibitem[\protect\citeauthoryear{Fernholz and Karatzas}{Fernholz and
  Karatzas}{2009}]{FK:2009}
Fernholz, R. and I.~Karatzas (2009).
\newblock Stochastic portfolio theory: an overview.
\newblock In A.~Bensoussan and Q.~Zhang (Eds.), {\em Mathematical Modelling and
  Numerical Methods in Finance: Special Volume, Handbook of Numerical
  Analysis}, Volume~XV, pp.\  89--168. Amsterdam: North-Holland.

\bibitem[\protect\citeauthoryear{Karatzas and Ruf}{Karatzas and
  Ruf}{2015}]{KR:2015}
Karatzas, I. and J.~Ruf (2015).
\newblock {Lyapunov} functions as portfolio generators.
\newblock Technical report, Columbia University and University College London.

\bibitem[\protect\citeauthoryear{Karatzas and Ruf}{Karatzas and
  Ruf}{2016}]{KR:2016}
Karatzas, I. and J.~Ruf (2016).
\newblock {Distribution of the time to explosion for one-dimensional
  diffusions}.
\newblock {\em Probability Theory and Related Fields\/}, {to appear}.

\bibitem[\protect\citeauthoryear{Karatzas and Shreve}{Karatzas and
  Shreve}{1991}]{Karatzas/shreve:1991}
Karatzas, I. and S.~E. Shreve (1991).
\newblock {\em {Brownian Motion and Stochastic Calculus}}.
\newblock New York: Springer-Verlag.

\bibitem[\protect\citeauthoryear{Karatzas and Shreve}{Karatzas and
  Shreve}{1998}]{Karatzas/Shreve:1998}
Karatzas, I. and S.~E. Shreve (1998).
\newblock {\em {Methods of Mathematical Finance}}.
\newblock New York: Springer-Verlag.

\end{thebibliography}
\end{document}